\documentclass{iopart}
\usepackage{epsfig}
\def\lsim{\hbox{ \raise.35ex\rlap{$<$}\lower.6ex\hbox{$\sim$}\ }}
\def\gsim{\hbox{ \raise.35ex\rlap{$>$}\lower.6ex\hbox{$\sim$}\ }}

\begin{document}
\title{A Note on the evolution of cosmic string/superstring networks}
\author{ M Sakellariadou} \address{Division of Astrophysics, Astronomy
and Mechanics, Department of Physics, University of Athens, GR-15784
Zografos (Athens), Hellas}
\ead{msakel@cc.uoa.gr}
\begin{abstract}
In the context of brane world scenario, cosmic superstrings can be
formed in D-brane annihilation at the end of the brane inflationary
era. The cosmic superstring network has a scaling solution and the
characteristic scale of the network is proportional to the
square root of the reconnection probability.
\end{abstract}


\maketitle

\section{Introduction}
Cosmic superstrings have recently gained a lot of interest. They are
expected to be formed in the context of a brane world scenario and it
was advertised that they lead to observational predictions which
might distinguish them from the gauge theory solitons, we are most
familiar with.

Cosmic strings are formed in a large number of Spontaneous
Symmetry Breaking (SSB) phase transitions. Suppose we have the
symmetry breaking of a group $G$ down to a sub-group $H$ of $G$.
The formation of cosmic strings is related to the first homotopy
group $\pi_1(G/H)$ of the vacuum manifold ${\cal M}={G/ H}$, i.e.,
if ${\cal M}$ is not simply connected then cosmic strings form. In
an exhaustive study of symmetry breaking schemes from a large
gauge group down to the standard model gauge group, Jeannerot
\etal~\cite{jrs} have shown that in the context of supersymmetric
grand unified theories, cosmic strings generically appear at the
end of a hybrid inflationary era. However, the measurements of the
Cosmic Microwave Background (CMB) temperature anisotropies impose
severe constraints on the cosmic strings contribution to the CMB
data, as it was shown by Bouchet \etal~\cite{bouchet}, and more
recently by Pogosian \etal~\cite{p1,p2}.  Thus, CMB measurements
can impose upper limits to the value of the dimensionless
parameter $G\mu\sim \eta^2/m_{\rm Pl}^2$, where the SSB scale
$\eta$ determines the critical temperature $T_{\rm cr}$ for the
phase transition leading to string formation, and $G=1/m_{\rm
Pl}^2$ is Newton's constant. Recently, Rocher and
Sakellariadou~\cite{jm1,jm2} have shown that F-term as well as
D-term supersymmetric hybrid inflation, accompanied by cosmic
string formation at the end of the inflationary era, are still
compatible with CMB measurements, provided the couplings and the
mass scales are tuned within acceptable limits.

As the string network evolves, a number of string intersections take
place, and this, with a given probability, can lead to an exchange of
partners. String intercommutations lead to the creation of kinks along
the long strings, which implies that long strings are not straight but
they develop wiggles, and thus they emit gravitational radiation, as
it was calculated by Sakellariadou~\cite{ms1}. In addition, the loops
which are chopped off from certain string intersections and
self-string intercommutations also emit gravitational
waves. Gravitational radiation is conventionally believed to be the
most efficient mechanism for string decay. This issue is not settled
since Vincent \etal~\cite{v1,v2} have shown that the string network
looses energy directly into scalar and gauge radiation. Numerical
studies by Vincent \etal~\cite{v1} have for the first time shown that
particle production, and not gravitational radiation, is the dominant
energy loss mechanism for cosmic string networks. Whether or not
gravitational radiation is the most efficient mechanism for cosmic
string decay is a very important issue, which however we will not
address here and we will come to this question later, in Ringeval and
Sakellariadou~\cite{rs}.

In the context of brane world scenario one naturally expects an
era of brane inflation, as proposed by Dvali and Tye~\cite{dt}.
The end of brane inflation is taken place when, either a brane and
an anti-brane, or a pair of branes oriented at angles, collide.
The inter-brane separation, which is an open string mode, plays
the r\^{o}le of the inflaton field, while the inflaton potential
emerges from the exchange of closed string modes between the
branes.  As the inter-brane separation becomes smaller than some
critical value, an open string mode stretching between the branes
becomes tachyonic and the rolling of the tachyon field describes
the decay of the pair of branes. This leads to heating of the
universe and the beginning of the big bang. Brane annihilation can
lead to the formation of lower dimensional D-branes\footnote{Due
to brane intersections leading to reconnection and unwinding,
Durrer \etal~\cite{rmm} argue that all D$p$-branes of dimension
$p>3$ disappear and that one of the stable  3-dimensional branes
plays the r\^ole of our Universe.} which are seen as topological
defects by brane observers. As it was shown by Sarangi and
Tye~\cite{st}, cosmic superstrings are copiously produced during
brane collisions.  These cosmic superstrings are D$p$-branes with
$(p-1)$ dimensions compactified. It is important that neither
monopoles, nor domain walls, nor textures, are produced during
this process.

Cosmic superstrings share a number of properties in common with the
familiar cosmic strings. It is however important to identify their
differences which may lead to distinctive observational signatures.

In what follows we examine the behavior of the characteristic scale of
the string network with respect to the string reconnection
probability. We reach conclusions which differ from results by Dvali
and Vilenkin~\cite{dv} and Jones \etal~\cite{jst}.  This may lead to
important consequences regarding the observational signatures of
cosmic superstrings and their differences with respect to their
solitonic analogues.

\section{Numerical experiments}
The picture we have in mind is brane annihilation leading to
independent stochastic networks of Dirichlet (D) and
Fundamental (F)
strings. In other words, we will not consider the interaction between
D- and F-networks. D- and F- strings may be observed as cosmic
strings. The evolution of string networks and thus, their
observational consequences, depend on the outcome of string
collisions. It is well known that reconnections may lead to closed
loop formation and the straightening of long open strings. Regarding
this issue, it seems to be an important difference between
superstrings and the gauge theory soliton strings.

The microscopic structure of the string core will only affect the
string evolution when two strings collide and their cores come to
an interaction. A basic difference between superstrings and cosmic
strings is that the first move in extra compact dimensions. This
motion leads to a suppression of the reconnection probability
since strings can miss each other.  The effect of extra dimensions
on superstring evolution can be seen by choosing a small
reconnection probability, as it was stated by Dvali and
Vilenkin~\cite{dv} and earlier it had been shown numerically in
the context of string gas cosmology by Sakellariadou~\cite{ms2}.
Performing numerical simulations of cosmic string evolution we
find that some aspects of the evolution of strings moving in
a 3-dimensional space with reconnection probability smaller than
1, are qualitatively comparable to that of strings moving in a
higher dimensional space with reconnection probability equal to 1.

For gauge theory solitons, the probability of reconnection when two
strings meet is equal to 1. As vortex strings in the Abelian Higgs
model collide, reconnection is classically inevitable, provided the
collision velocity is small and the relative angle between the
orientation of the motion of the colliding strings is also small, as
shown by Hanany and Hashimoto~\cite{hh1}.

The collisions
between all possible pairs of superstrings have been studied in string
perturbation theory by Jackson \etal~\cite{jjp}. For F-strings, the
reconnection probability is of order of $g_{\rm s}^2$, where $g_{\rm
s}$ stands for the string coupling.  For F-F string collisions, it was
found by Jackson \etal~\cite{jjp} that the reconnection probability
$P$ is $10^{-3}\lsim P\lsim 1$. For D-D string collisions, one has
$10^{-1}\lsim P\lsim 1$. Finally, for F-D string collisions, the
reconnection probability can take any value between 0 and 1.

The results by Jackson \etal~\cite{jjp} have been confirmed by a
quantum calculation of the reconnection probability for colliding
D-strings. More precisely, classically nothing happens for colliding
D-strings, since the D-string action permits a solution describing
D-strings passing through each other.  One has to study colliding
D-strings quantum mechanically, and then one gets a a tachyonic
instability, which is characteristic for intersecting D-branes.  The
tachyon condensation leads then to D-string reconnection, with
reconnection probability derived from the time evolution of the
tachyon wave function~\cite{hh1}.

Following Sen's conjecture~\cite{sen}, we consider the D-strings as
vortex-like topological defects in tachyon condensation on a brane and
anti-brane pair.  We study numerically the evolution of the
superstring network as a function of the reconnection probability. Our
initial string configuration is generated by the Monte Carlo algorithm
of string formation, given by Vachaspati and
Vilenkin~\cite{vv}. Alternatively, either we artificially form an
initial string network consisting of long winding strings accompanied
by a loop gas, or we just consider a gas of small loops. This approach
is often preferable since we have control over the initial string
density. The strings are represented by discrete points on a lattice
and they are contained in a d-dimensional box with periodic boundary
conditions, i.e. a torus. This implies that all strings are closed
loops. Our results are independent of the choice of boundary
conditions and we have checked their validity by choosing instead
reflecting boundary conditions. Points on strings are equally spaced
in energy, and therefore the total energy of a loop is proportional to
the number of points by which it is discretised. We choose the energy
between neighboring points on a string as being the unit of energy and
the lattice spacing as the unit of length and time.  Strings formed
with a Monte Carlo algorithm are initially at rest and are composed of
straight links pointing in three orthogonal directions. When the
initial string network is composed from a loop gas, the strings have
initially non-zero velocities. The string network is divided into long
strings (strings of {\sl infinite} length) and small loops. We have checked
that this division is not sensitive to the exact choice of the
separating line, since we find only very few loops in the intermediate
range of sizes.

In our numerical simulations we have tested the robustness of our
findings against three possible definitions of {\sl infinitely} long
strings. (1) We define an {\sl infinitely} long string, a loop whose energy
is equal or greater to the square of the size (or half the size) of
the box.  This choice is made since on large scales the strings are
Brownian~\cite{svtherm}.  (2) We define an {\sl infinitely} long string, a closed loop for
which at least one of its extents exceeds $k\xi$, where $k\sim 1$ is a
numerical coefficient and the characteristic length scale $\xi(t)$ is
determined using the density of long strings from the previous
evolution timestep. An alternative definition using the extents, is to
say that a string is {\sl infinitely} long if at least one of its extents is
greater than the size (or half the size) of the box. (3) We define a
string as being {\sl infinitely} long if it satisfies both criteria (1) and
(2). Our results are not sensitive to the choice of one of the
conditions (1), (2) or (3).

We let the string network evolve by discretising the Nambu equations
of motion in a Minkowski space. When two strings intersect, they
intercommute with a given probability $P$. To approach the realistic
expanding universe, we do not allow small loops to interact with the
rest of the network.  Thus, small loops are not allowed to reconnect
to long strings at later stages of the evolution; they freeze out.  We
decouple a small loop from the string network if all its extents are
smaller than $k'\xi$, where $k'$ is an adjustable numerical factor
$k'\sim 1$. We let the string network evolve up to $L/2$ timesteps
since late-time behavior of the network is dominated by finite-size
effects.

The long strings are
characterised by a single length scale
\begin{equation}
\xi(t)=\left({\rho_{\rm l}\over \mu}\right)^{-1/2}~,
\end{equation}
where $\rho_{\rm l}$ denotes the energy density of long strings and
$\mu$ stands for the linear mass density, which equals the string
tension.  Thus, the typical distance between the nearest string
segments and the typical curvature of the strings are both of
order $\xi$.  We study the characteristic length $\xi$ as a function
of evolution time $t$, performing a number of runs with a range of
values for the free parameters of the model.  These free parameters
set, for a box of a given size $L$, the reconnection probability $P$,
the energy $E_{\rm c}$ of the smallest allowed loop, the definition of
an {\sl infinitely} long loop (i.e., $k$), and the definition of the very
small loops which freeze out in the subsequent evolution (i.e., $k'$).
Most of our numerical experiments were done in a 3-dimensional box of
size $L=60, 80, 120, 160, 200$ with parameters $\xi(t=0)=4, 6$;
$E_{\rm c}=2, 4, 6$; $k=0.75, 1.5, 2$; $k'=0.5, 0.7, 1$ and $P\in
[10^{-4},~ 1]$.  We find that $\xi$ grows linearly with time.
\begin{equation}
 \xi(t)\propto \zeta t ~,
\end{equation}
where the slope $\zeta$ depends on the reconnection probability
$P$, as well as on the energy of the smallest allowed loops (energy
cut-off).

We now turn to the dependence of the slope $\zeta$ on the reconnection
probability $P$. This will set the dependence of the length scale
$\xi$ on the reconnection probability $P$. Our results are shown in
Fig.~1 below.

\begin{figure}[hhh]
\begin{center}\includegraphics[scale=.8]{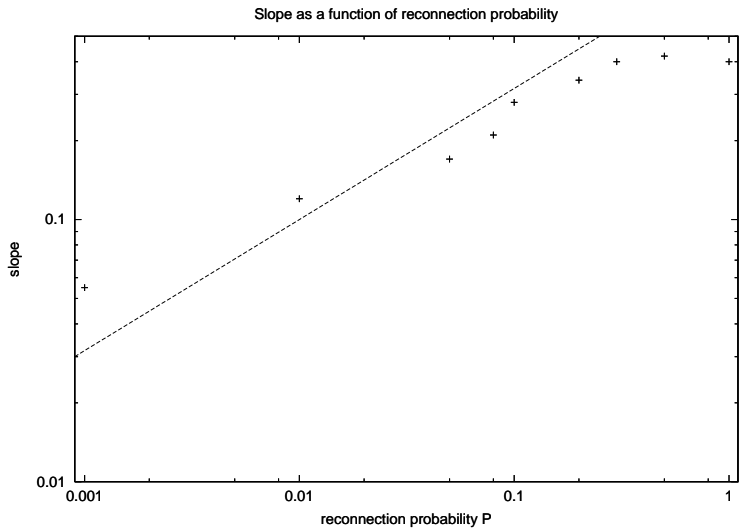}
\caption{The slope $\zeta$ as a function of the reconnection
  probability $P$ in a logarithmic scale.
The dots give our data.  The straight line shows the square-root fit.    }
\end{center}
\end{figure}

Clearly, for small values of the reconnection probability, a good fitting
is
\begin{equation}
\zeta \propto \sqrt{P} \Rightarrow \xi(t)\propto \sqrt{P} t~,
\label{law}
\end{equation}
for reconnection (or intercommuting) probability in the range
$0.001\lsim P \lsim 0.3$. We thus agree with the old results by
Sakellariadou and Vilenkin~\cite{sv}, while we do not agree with
the more recent statements by Dvali and Vilenkin~\cite{dv}, and
Jones \etal~\cite{jst}.  More precisely, Dvali and
Vilenkin~\cite{dv} claim that for relativistic strings the scale
of the network should be proportional to the reconnection
probability and to the time evolution, namely $\xi\sim P t$. The
explanation of our square root dependence on the reconnection
probability was suggested to have an origin on the presence of
small-wiggles superimposed on the long strings. These wiggles
could lead to a higher number of reconnection opportunities per
string encounter. To check this possible explanation we have
artificially created large straight strings winding around a cubic
box. For such a network we have studied the dependence of the
characteristic scale on the reconnection probability, for a range
of values for $P$ lower than the one considered earlier in the
numerical study by Sakellariadou and Vilenkin~\cite{sv}. Our
current studies confirm the numerical fit to the data given in
Eq.~\ref{law}.

In the same way, our results are in disagreement with
statements by Jones \etal~\cite{jst}.  It is clear that as a long
string of length $l$ moves with velocity $u$ sweeps out a surface
equal to $l u(\Delta t) / D^2$ in a unit time interval $\Delta t$,
where $D$ denotes the size of the box. We set $l=D\equiv L$. Then,
the number of collisions per unit time interval $\Delta t$,
between two long strings is $(u/L)N$, where $N$ stands for the
number of long strings in the whole volume.  Thus, the number of
intercommutations for relativistic strings per unit time and per
unit volume is $PN(1/L)(1/L^3)$, where $P$ stands for the
intercommutation probability per string intersection. Up to this
point, we basically agree with Jones \etal~\cite{jst}. However,
the misleading point is that string intersections do not, in
principle, chop off loops. Thus intersections between two long
strings is not the most efficient mechanism for energy loss of the
string network. The possible string intersections can be divided
into three possible cases, which are drawn in Figs.~2, 3, 4
respectively.  These cases are: (1) Two long strings collide in
one point and exchange partners with intercommuting probability
$P_1$; (2) two strings collide in two points and exchange partners
chopping off a small loop with intercommuting probability $P_1^2$;
and (3) one long string self-intersects in one point and chops off
a loop with intercommuting probability $P_2$, which in general is
different than $P_1$.

\begin{figure}[hhh]
\begin{center}\includegraphics[scale=.6]{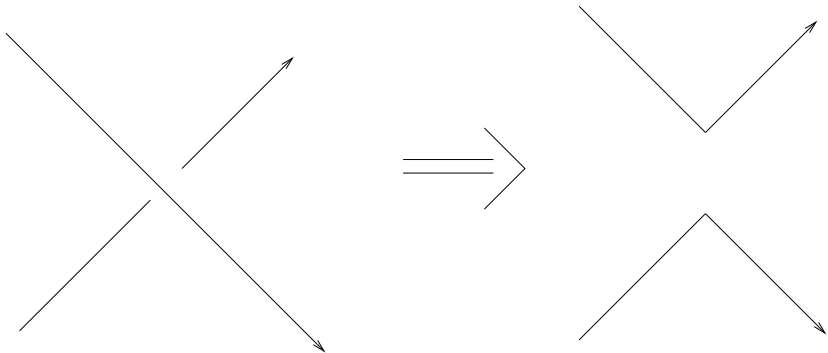}
\caption{Illustration of string-string intersections in one
  point, and the formation of two new long strings via exchange of partners.}
\end{center}
\end{figure}

\begin{figure}[hhh]
\begin{center}\includegraphics[scale=.5]{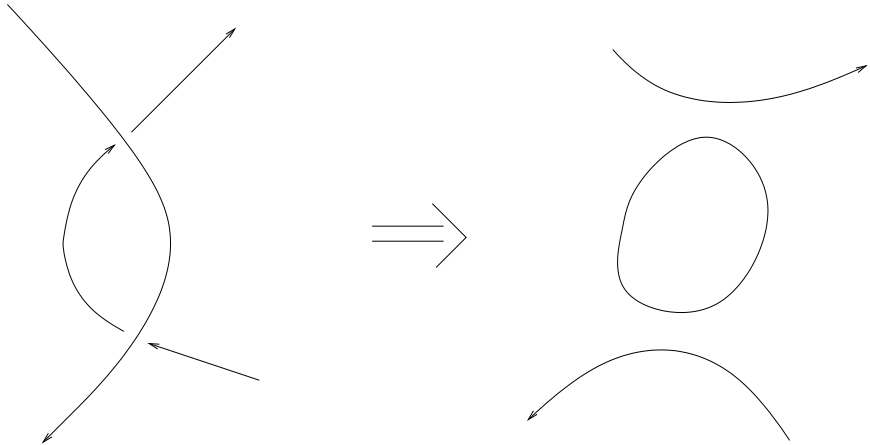}
\caption{Illustration of string-string intersections in two
  points. Two new long strings are created as well as a closed loop.}
\end{center}
\end{figure}

\begin{figure}[hhh]
\begin{center}\includegraphics[scale=.5]{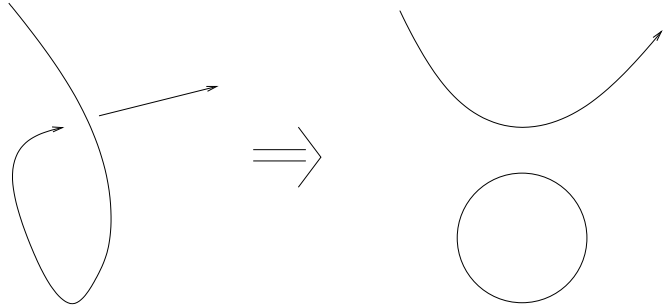}
\caption{Illustration of self-string intersections leading to
the formation of one new long string and a closed loop.}
\end{center}
\end{figure}

Clearly, only cases (2) and (3) lead to a closed loop formation and
therefore remove energy from the long string network.  Between cases
(2) and (3), only case (3) is an efficient way of forming loops and
therefore dissipating energy.  We have checked numerically that case
(3) appears more often than case (2), and besides, case (2) has in
general a smaller probability, since one expects that $P_1\sim P_2$.
However, the heuristic argument employed by Jones \etal~\cite{jst}
does not refer to self-string intersections (i.e, case (3)); it only
applies for intersections between two long strings. This is clear
since intersections between two long strings depend on the string
velocity, however self-string intersections should not depend on how
fast the string moves. In other words, a string can intersect itself
even if it does not move but it just oscillates
locally\footnote{Recently, Damour and Vilenkin~\cite{dvgw} also agree
with the square-root fit.}.

A cosmic string network has at formation $80\%$ of its energy
density in the form of long strings and only $20\%$ in closed
string loops~\cite{vv}. A thermodynamical study of the string
network has shown that as we increase the string energy density,
we reach a critical value, the Hagedorn energy density, which is
characterised by the appearance of long strings~\cite{svtherm}.
Above the critical energy density, whose value depends on the
dimensionality of the space where the strings move~\cite{ms2}, the
extra energy density goes to long strings, while below the
critical energy density the string network chops off small closed
loops and the long strings disappear, provided the space
dimensionality is $d=3$ and the reconnection probability $P=1$. In
the low energy regime, closed string loops are constantly produced
as long strings intersect or self-intersect, provided the
reconnection probability is large. As $P$ increases, the number of
string intersections increases, leading to a decrease of long
strings and an increase of small closed loops. Thus, the energy
density of long strings is inversely proportional to the
reconnection probability $P$, while the energy density of closed
string loops is proportional to $P$.

Performing numerical simulations, we address the question of scaling
of the long strings network. More precisely, for $P\in [10^{-4},1]$,
we study the time evolution of the slope $\zeta$. We find that $\zeta$
reaches a constant value at relatively the same time $t$ for various
values of $P$. This implies that the long strings reach
scaling\footnote {We would like to bring to the attention of the
reader that Avgoustidis and Shellard~\cite{as04} reached different
conclusions regarding the scaling issue.}.  We illustrate our results
in Figs.~5,~6 below.

\begin{figure}[hhh]
\begin{center}
\includegraphics[scale=.35]{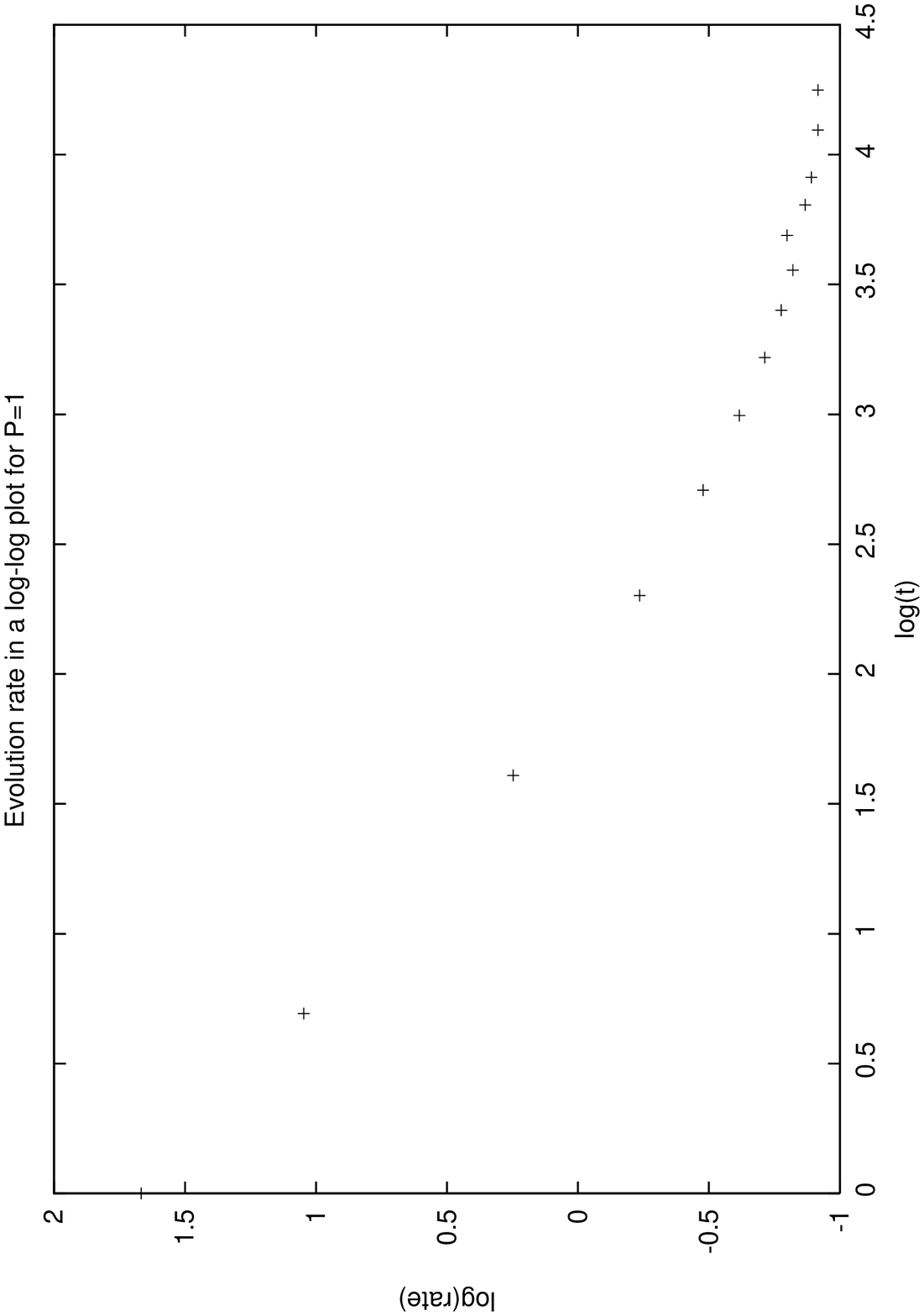}
\includegraphics[scale=.35]{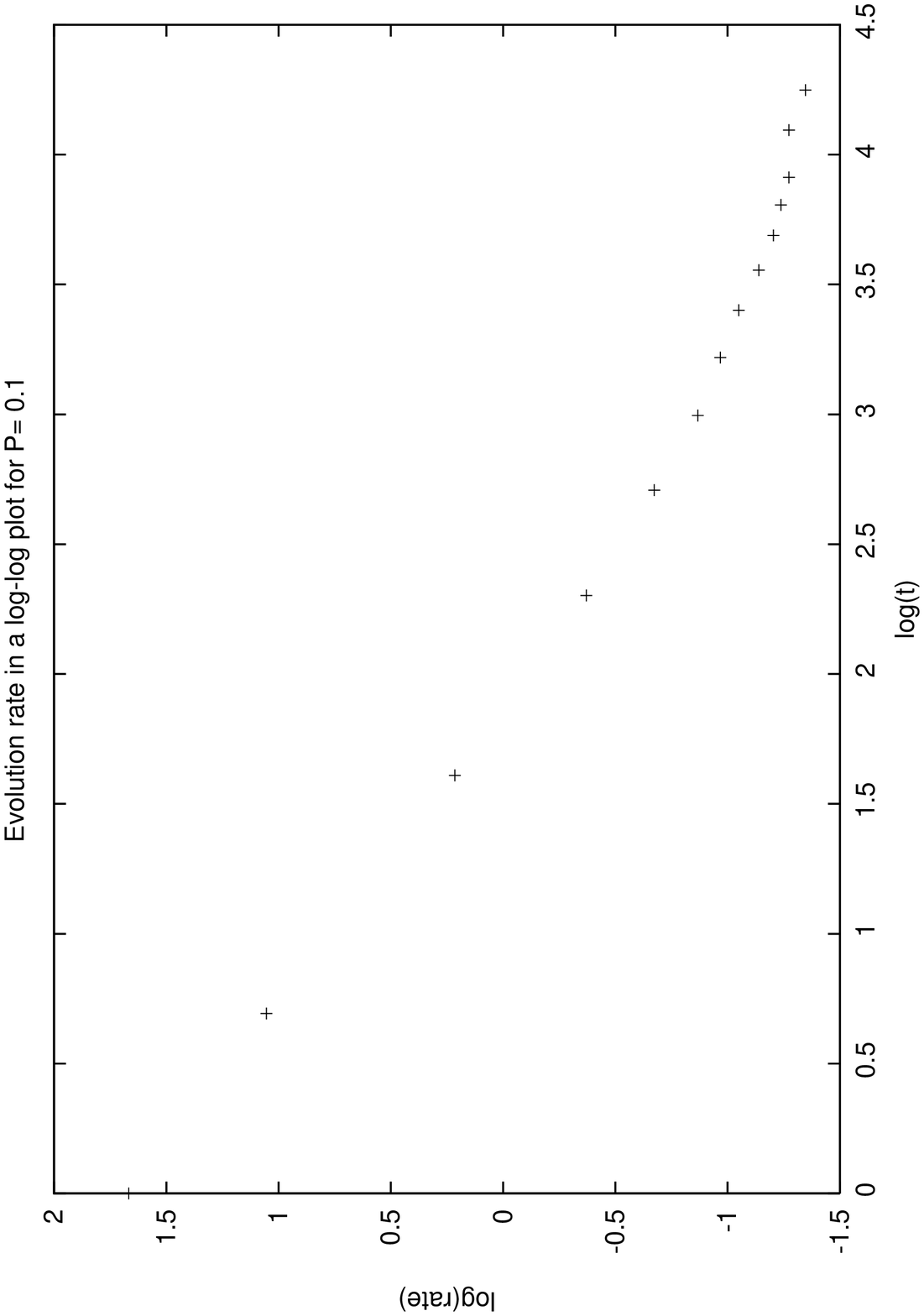}
\caption{Time evolution of the rate $\zeta$ in a log-log plot. The
strings are moving in a cubic box of size $L=80$ with $N=12996$
lattice points occupied by strings. The reconnection probability is
$P=1$ for the left panel and $P=0.1$ for the right one.}
\end{center}
\end{figure}

\begin{figure}[hhh]
\begin{center}
\includegraphics[scale=.35]{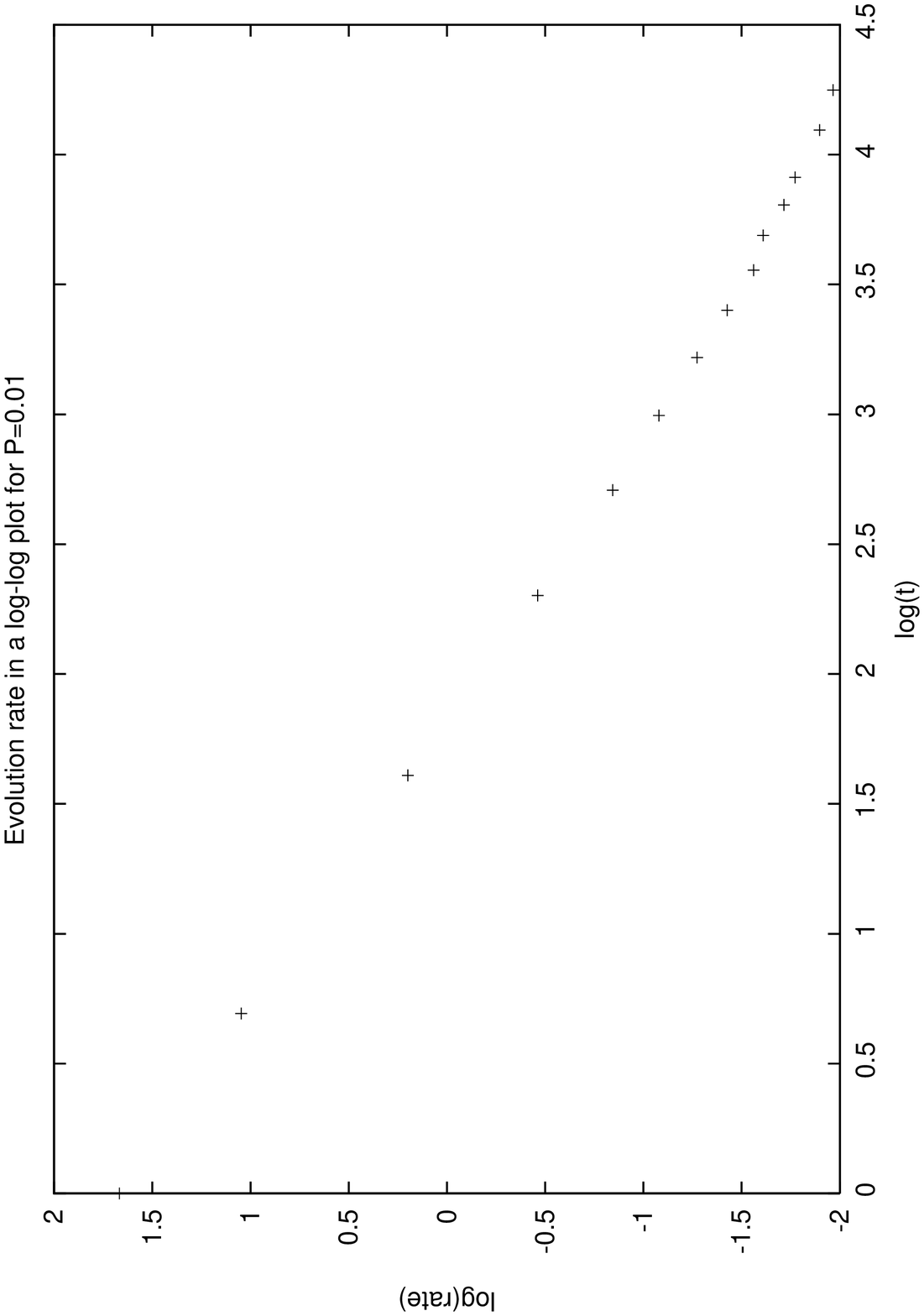}
\includegraphics[scale=.35]{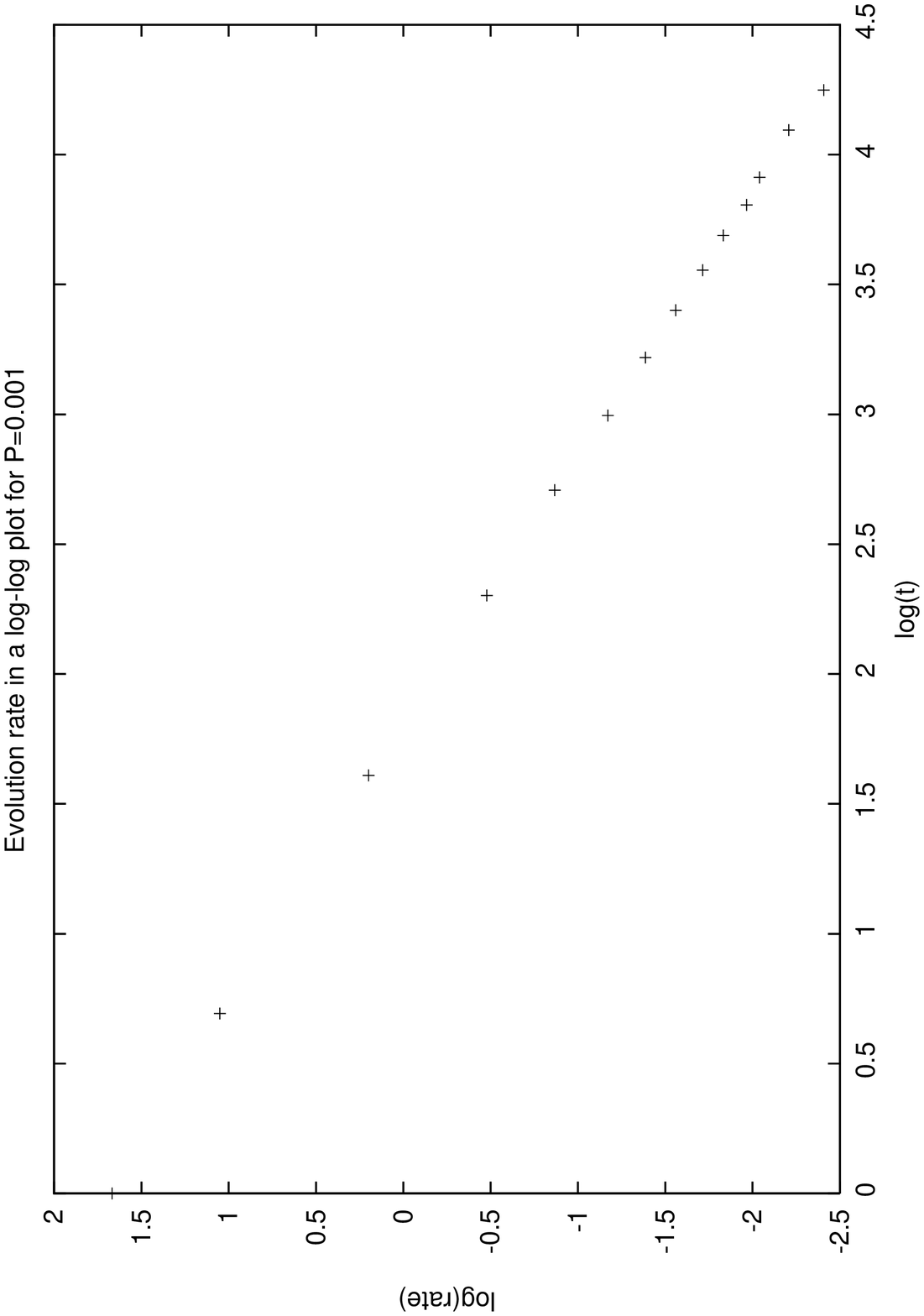}
\caption{Time evolution of the rate $\zeta$ in a log-log plot. The
strings are moving in a cubic box of size $L=80$ with $N=12996$
lattice points occupied by strings. The reconnection probability is
$P=10^{-2}$ for the left panel and $P=10^{-3}$ for the right one.}
\end{center}
\end{figure}

Our network is characterised by two components: (1) a few long strings
with a scale-invariant evolution; the characteristic curvature radius
of long strings, as well as the typical separation between two long
strings are both comparable to the horizon size, $\xi(t)\simeq {\sqrt P}
t$, and (2) a large number of small closed loops having sizes $\ll t$.

We have studied the energy distribution of small string loops in the
high energy density regime. For strings moving in a 3-dimensional
lattice, the number of small loops, with a given energy, per unit
volume is shown in Figs.~7,~8 below for several values of the
reconnection probability $P$. In these figures we only change the
value of the reconnection probability and we see that the results seem
not to be sensitive on the exact value of P. Clearly, as the
reconnection probability decreases, the relaxation time increases.
Thus, to get the energy distribution of short loops we have to wait
much longer time, if the reconnection probability is well below 1. The
energy distribution of short loops, with a given energy, per unit
volume is well fitted by
\begin{equation}
\label{en-dis-1}
{dn\over dE}\sim E^{-5/2} ~~~~{\rm for }~~~~ \rho>\rho_{\rm H}(=0.172)~.
\end{equation}

\begin{figure}[hhh]
\begin{center}
\includegraphics[scale=.35]{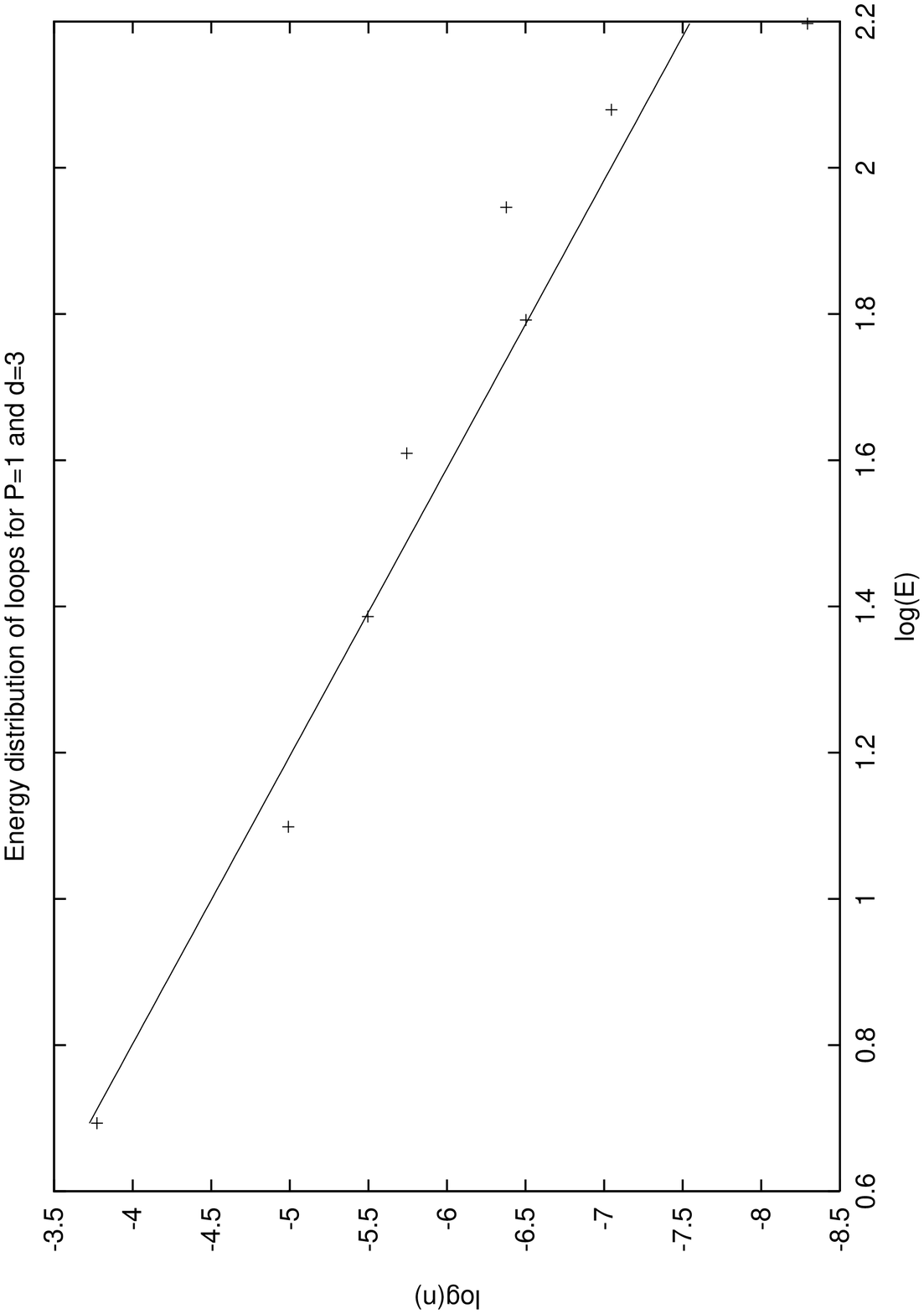}
\includegraphics[scale=.35]{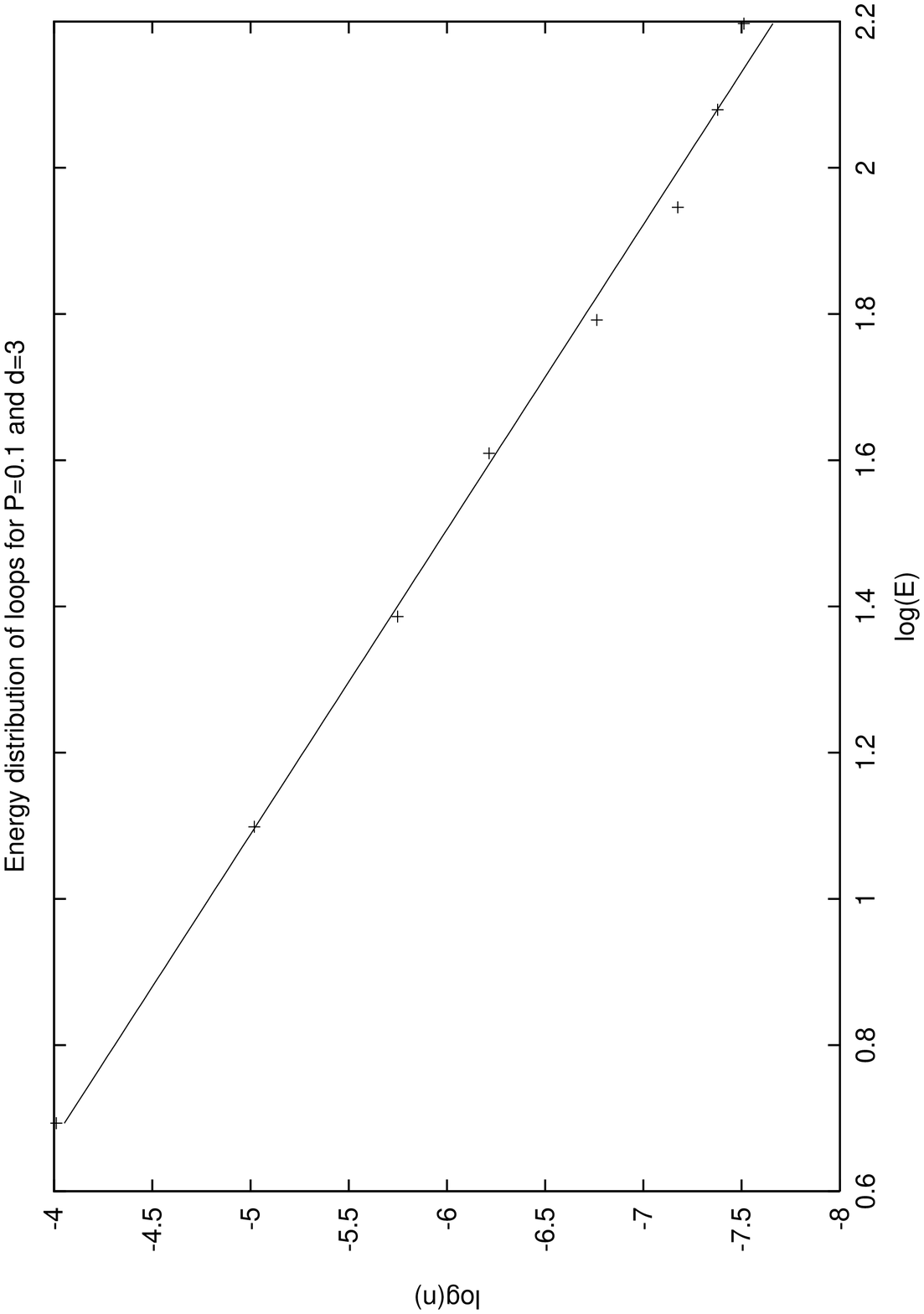}
\caption{Energy distribution of small string loops in the high-energy
regime. The strings are moving in a 3-dimensional torus and the
reconnection probability is equal to $P=1$ for the left panel and
$P=0.1$ for the right one. This distribution does not depend on the
particular value of the string energy density $\rho$, provided
$\rho>\rho_{\rm H}=0.172$~\cite{ms2}, and is well fitted by $dn/dE\sim
E^{-5/2}$, which is also plotted.}
\end{center}
\end{figure}

\begin{figure}[hhh]
\begin{center}
\includegraphics[scale=.35]{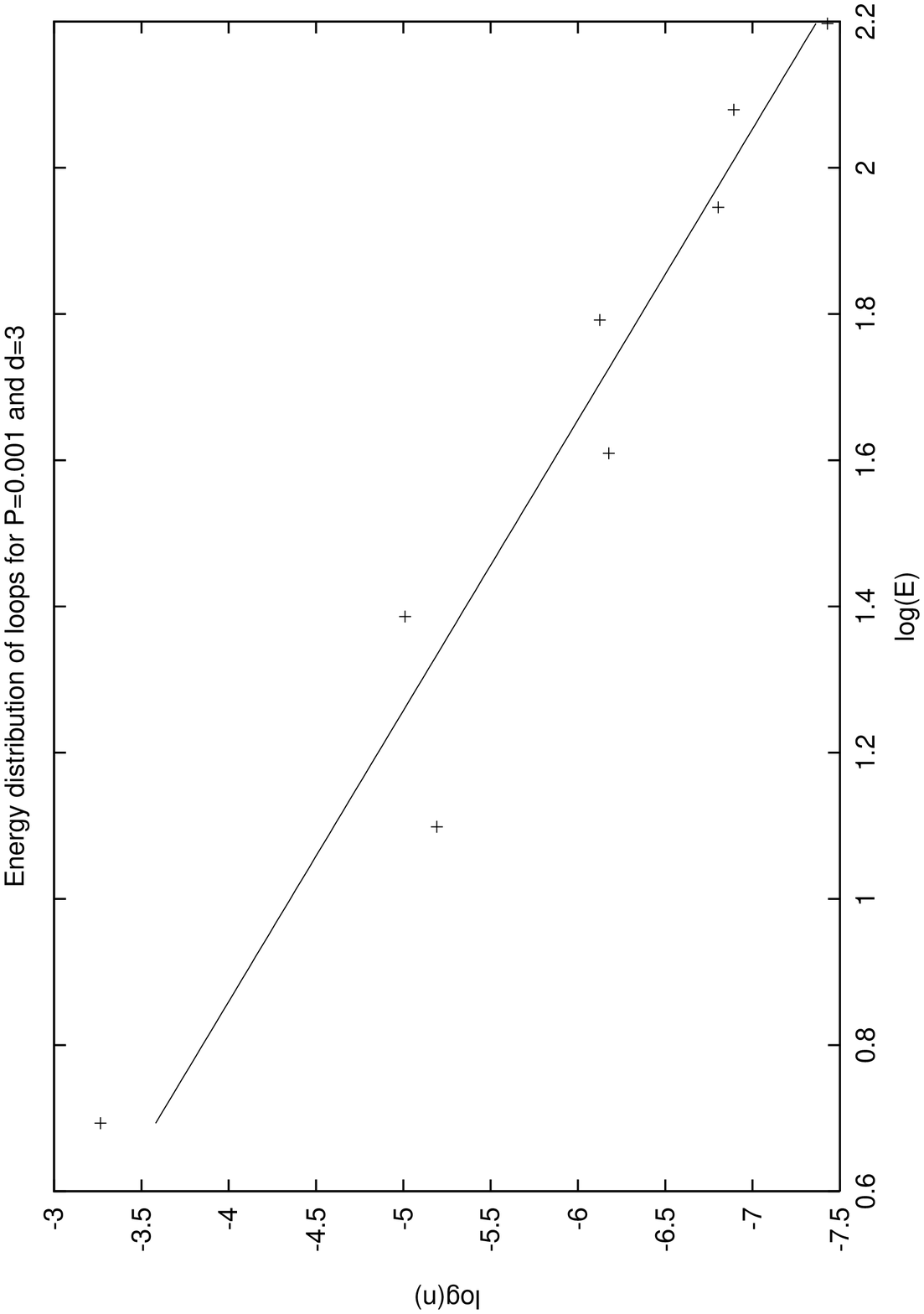}
\caption{Energy distribution of small string loops in the high-energy
regime. The strings are moving in a 3-dimensional torus and the
reconnection probability is equal to $P=0.001$.  This distribution
does not depend on the particular value of the string energy density
$\rho$, provided $\rho>\rho_{\rm H}=0.172$~\cite{ms2}, and is well
fitted by $dn/dE\sim E^{-5/2}$, which is also plotted.}
\end{center}
\end{figure}

We then perform the same study but for strings moving in a higher
dimensional space with reconnection probability $P=1$. In Fig.~9 below
we show the energy distribution of short loops, in the high energy density
regime, for strings moving in a 4-dimensional torus with reconnection
probability $P=1$. We see that the number distribution is well fitted
by
\begin{equation}
\label{en-dis-2}
{dn\over dE}\sim E^{-3} ~~~~{\rm for }~~~~ \rho>\rho_{\rm H}(=0.062)~.
\end{equation}
Thus, Eqs.~(\ref{en-dis-1}),~(\ref{en-dis-2}) confirm the behavior
\begin{equation}
\label{en-dis-l}
{dn\over dE}\sim E^{-(1+d/2)}~,
\end{equation}
where $d$ stands for the space dimensionality of the torus in
which the strings move.

\begin{figure}[hhh]
\begin{center}
\includegraphics[scale=.35]{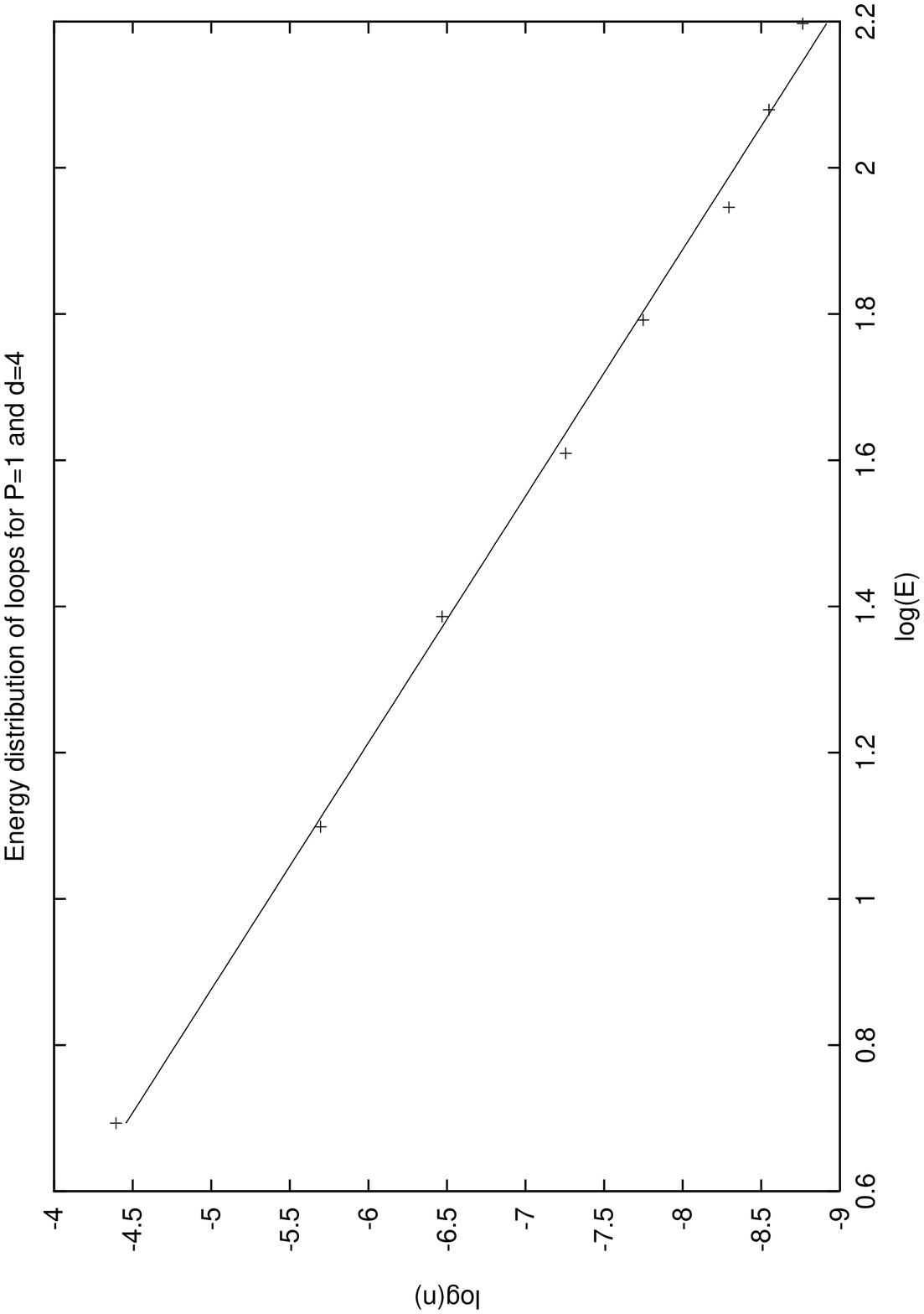}
\caption{Energy distribution of small string loops in the high-energy
regime. The strings are moving in a 4-dimensional torus and the
reconnection probability is equal to $P=1$.  This distribution does
not depend on the particular value of the string energy density
$\rho$, provided $\rho>\rho_{\rm H}=0.062$~\cite{ms2}, and is well
fitted by $dn/dE\sim E^{-3}$, which is also plotted.}
\end{center}
\end{figure}

Our results lead to the following asymptotic energy density of the
long strings network, providing there are string interactions (i.e.,
$P\neq 0$):
\begin{equation}
\rho_{\rm l}={\mu\over P t^2}~.
\end{equation}
This implies that the extra dimensions increase the energy density
of the long strings, which however results to a less enhancement
than what it was previously believed.  As a result, for fixed
linear mass density, the cosmic superstring energy density may be
higher than the field theory case, but at most only by one order
of magnitude. More precisely, the fraction of the total density in
the form of strings in the radiation-dominated era reads
\begin{equation}
{\rho_{\rm str}\over \rho_{\rm total}}={32\pi\over 3} {G\mu\over P}~.
\end{equation}

Oscillating string loops loose energy by emitting graviton, dilaton
and Ramond-Ramond (RR) fields. These three mechanisms for energy loss
are comparable to each other provided the dilaton and RR fields can be
considered as massless. Constraints on the allowed energy scale of
cosmic strings from the observational bounds on dilaton decays have
been earlier discussed by Damour and Vilenkin~\cite{tdv}. In the case
of cosmic superstrings, the effect of the extra compact dimensions on
the superstring evolution will decrease the number density of the
chopped off closed loops. Since the energy density of long strings is
inversely proportional to the reconnection probability $P$, provided
string interactions are present in the network, the energy density of
closed loops is proportional to $P$. Clearly the number density of
these loops at the time of decay is also proportional to $P$. Thus,
the dilaton density parameter $Y_\phi=n_\phi(t)/s(t)$, where
$n_\phi(t)$ denotes the dilaton density and $s(t)$ the entropy
density, should be proportional to the reconnection probability $P$.
Assuming that the mass of the dilaton field is $m_\phi \sim 1 {\rm
TeV}$, Damour and Vilenkin~\cite{tdv} have set upper bounds to
$Y_\phi$, which depend on the dilaton lifetimes $\tau$. For dilaton
lifetime in the range $10^7{\rm s}\lsim \tau\lsim t_{\rm dec}$, one
roughly has $Y_{\phi}\lsim 1.4 \times 10^{-12} (m_\phi/ 1{\rm
GeV})^{-1}$, where for a dilaton mass $m_\phi\sim 1{\rm TeV}$ we
obtain $G\mu\lsim P^{-2/3} 10^{-16}$, which leads to an upper bound
$\eta\lsim P^{-1/3} 10^{11}{\rm GeV}$ for the energy scale of cosmic
superstrings, which determines the critical temperature for the
transition. Thus, a lower reconnection probability allows a higher
energy scale of strings, by at most one order of magnitude.

The cosmic superstring network formed towards the end of the brane
inflationary era, can lead to a number of observational predictions,
which in principle can be distinguishable from the signatures of gauge
theory solitons. Thus, cosmic superstrings might be observable, at
least in principle, via gravitational lensing, linear discontinuities
on the angular spectrum of the CMB temperature anisotropies, or
through the spectrum of gravity waves~\cite{dvgw}. Clearly, these signatures
depend on the reconnection probability $P$. Since $P$ for cosmic
superstrings is lower than in the case of gauge theory solitons,
the evolution of the superstring network will slow down, in the
sense that the network will have a larger number of long strings
with somehow less kinks than in the case of cosmic string networks
where $P=1$. Similarly, the process of closed loop formation slows
down in the case of cosmic superstrings.

\section{Conclusions}
Cosmic superstrings are expected to be the outcome of brane inflation
in the context of a brane world scenario. These objects are quite
similar to their solitonic analogues, even though they have some
distinguishable features which may lead to different observable
signatures. The main difference between cosmic superstrings and cosmic
strings is the fact that the former live in extra dimensions. This
results to a smaller reconnection probability, which will alter the
subsequent evolution of a superstring network as compared to the most
familiar string network evolution.

We have studied numerically the evolution of a cosmic superstring network
 in a Minkowski space. We have found
that the network reaches a scaling solution, which means that the
characteristic length scale of the string network, which gives the
typical curvature radius of long strings and the characteristic
distance between long strings, grow both proportional to the
Hubble radius $1/H$, or the age of the universe $t$.  In addition,
we found that the evolution rate $\zeta$ is proportional to the
square root of the reconnection probability. This will effect the
string energy density and may lead to distinguishable
observational signatures.

\end{document}